# High-power, high repetition rate, tunable, ultrafast vortex beam in the near-infrared

A. Aadhi,* and G. K. Samanta

Photonic Sciences Lab, Physical Research Laboratory, Navarangpura, Ahmedabad 380009, Gujarat, India

*Corresponding author: aadhi@prl.res.in

We report on experimental demonstration of high power, ultrafast, high repetition rate vortex beam source tunable in the near-IR wavelength range. Based on single-pass optical parametric generation of Yb-fiber laser of vortex order $l_p=1$ in a 50 mm long MgO doped periodically poled $LiNbO_3$ crystal, the source produces signal beam in vortex profile of order $l_s=1$ across 1433–1553 nm. Additionally, the source produces broadband idler radiation tunable across 3379–4132 nm in the Gaussian beam profile. We observed that the vortex profile of the pump beam is always transferred to the signal beam due to the highest overlapping integral among the interacting beams and the idler maintains a Gaussian spatial profile owing to conservation of orbital angular momentum in optical parametric processes. For pump power of 4.72 W, the signal and idler beams have maximum power of 1.7 W at 1509 nm and 0.48 W at 3625 nm respectively. The signal vortex beam has output pulses of width ~637 *fs* at a repetition rate of 78 MHz. The signal (idler) has spectral width of 4.3 nm (129.5 nm) and passive peak-to-peak power fluctuation better than 3% (1.1%) over 30 minutes, respectively.

Light beams having screw-like phase dislocation is known as optical vortices. The doughnut intensity distribution and the existence[1] of OAM make the optical vortices indispensable for the variety of applications in science and technology including quantum information[2], high-resolution microscopy[3], particle micromanipulation and lithography[4], and material processing[5]. Typically, optical vortices are generated through the mode conversion of optical beam of Gaussian spatial profile. The most common mode converters include computer generated holography technique based on spatial light modulators (SLMs)[6], spiral phase plates (SPPs)[7], cylindrical lenses[8] and the spiral MEMS mirrors[9]. However, all these converters suffer from at least one of the common drawbacks such as low power handling capabilities, limited wavelength coverage and high costs. Therefore, it is imperative to explore alternative techniques to generate high power optical vortices tunable over wide wavelength range. Nonlinear frequency conversion has proven to be the most direct route to access various spectral domains across electromagnetic spectrum[10]. Therefore, efforts have been made previously to generate high power and higher order optical vortices through second-harmonic-generation (SHG), third-harmonic-generation (THG), optical parametric amplifications (OPA) and optical parametric oscillators (OPOs)[11-14]. Although the OPOs, as compared to SHG and THG, can provide optical vortices across wide wavelength range from a single device, the inherent complexity of the OPOs have restricted their use especially for the generation of vortex beams.

Alternatively, single-pass optical parametric generation (OPG) provides tunable optical radiation in a simple and compact system configuration. Therefore, the OPG of an optical vortex beam can be the most straightforward technique for generating tunable optical vortex. Unfortunately, due to low parametric gain, OPG are commonly performed with high pump intensities through the use of high energy, low repetition rate (RR) lasers along with suitable pump focusing and crystal length[15]. In contrary to the use of thin nonlinear crystal to accommodate broad spectrum of the ultrafast lasers, recently, we have demonstrated the possibility of single-pass OPG of a high RR femtosecond laser in Gaussian spatial profile in a long nonlinear crystal[16]. Using the extended phase-matching[17], where the spectral acceptance bandwidth is inversely proportional to the crystal length, we observed that a 50-mm long MgO-doped periodically poled lithium niobate (MgO:PPLN) crystal has an OPG threshold of 1.24 W for an ultrafast laser having output pulses of width ~260 fs at 78 MHz in Gaussian spatial beam profile[16]. Such a low value of OPG threshold opens up the possibility of generating tunable optical beams in different spatial structure through single-pass OPG process.

Here, we report on single-pass OPG of vortex pump beam producing high power and high RR ultrafast tunable optical radiation in vortex spatial profiles. Pumped with a vortex beam of order, $l$=1, the OPG produces signal beam continuously tunable across 1433-1553 nm with power as high as 1.7 W at 1509 nm in vortex beam profile and the corresponding idler in Gaussian profile across 3379-4132 nm with a maximum idler power of 0.48 W at 3625 nm for 4.66 W of pump power.

The schematic of the experimental setup is shown in Fig. 1. A 5-W average power, Yb-fiber laser (Fianium, FP1060-5-*fs*) producing femtosecond pulses of temporal width (full-width at half maximum, FWHM) of 260 *fs* in *sech²* pulse shape at a repetition rate of 78 MHz is used as the pump source for OPG process. The laser has spectral-width (FWHM) of ~15 nm centered at 1064 nm. A polarizer based power attenuator[18] is used to vary the pump power into the nonlinear crystal. The pump beam is transformed into optical vortex of order, $l_p=\pm 1$, by using a spiral phase plate (SPP) having thickness variation in the transverse plane corresponding to the phase winding number, $l$=1, at 1064 nm. The sign of the vortex can be changed by simply flipping the SPP. A 50-mm long MgO:PPLN crystal, with grating period varying in the range $\Lambda$=28.5–31.5 μm in 0.5 μm step, is used for the OPG process. The crystal housed in an oven whose temperature can be varied up to 200°C with a stability of ±0.1°C. The half-wave plate ($\lambda/2$) adjusts the polarization of the pump beam to the crystal. The lens of focal length, $f$=100 mm, is used to focus the pump beam at the center of the nonlinear crystal. Two wavelength separators, $S_1$ and $S_2$ extract signal and idler radiations from the residual pump beam. The power as mentioned throughout the manuscript, except where otherwise noted, represents the average power of the optical beams.

To verify the generation of vortex beam, we operated the MgO:PPLN crystal near to zero GVM region[16,17] by setting the crystal temperature and grating period, T=100°C and $\Lambda$=30 μm respectively. Under this operating condition, the signal and idler wavelengths are measured to be 1523 nm and 3572 nm. Pumping the crystal with a constant power of $P_p$=4.66 W and vortex order of, $l_p$=±1, we measured the intensity distribution of the signal and idler using a pyroelectric beam profiler with the results shown in Fig. 2. For the pump beam of vortex order, $l_p$=1 (Fig. 2a), the signal beam carries the doughnut intensity distribution (Fig. 2b) whereas the idler beam has a Gaussian intensity distribution (Fig. 2c). To confirm the vortex order, we have analyzed pump, signal and idler beams using tilted lens technique[17] with the results shown in second column of Fig. 2. Given that the vortex beam of order, $l$, after passing through a tilted lens split into $|l|+1$ numbers of bright lobes at the focal plane[19], the images of second column, (d-f) of Fig. 2, confirm the vortex order of the pump, signal and idler as $l_p$=1, $l_s$=1 and $l_i$=0 (Gaussian) respectively, and the OAM conservation in OPG process, $l_p$=$l_s$+$l_i$. Similarly, from the third column, (g-i), showing the intensity distribution of pump, signal, and idler beams respectively, it is evident that for pump beam having vortex order, $l_p$=-1 the signal and idler carry doughnut and Gaussian intensity distribution respectively. Splitting of the beams using tilted lens as shown by fourth column, (j-l), show that both pump and signal have vortex order, $l_p$=$l_s$=-1, and the idler has vortex order of $l_i$=0 (Gaussian). The sign (±), a signature of the direction of phase variation, of the vortex beams are evident from the orientation of the bright spots as shown by the images in second and fourth columns of Fig. 2. According to the OAM conservation in parametric processes[13], the signal and idler beams have equal probability to carry the OAM of the pump beam, however, in the present experiment, it is also interesting to note that the pump vortex is directly transferred to the signal whereas the idler maintains a Gaussian intensity distribution.

To understand the OAM transfer mechanism in OPG process, we have evaluated the spatial overlap integral of pump, signal, and idler radiations along the crystal[20]. As reported previously[13,20], the normalized overlap integral, $\eta_{l_p,l_s,l_i}$, have nonzero value for $l_p$=$l_s$+ $l_i$. Therefore, for a fixed pump OAM mode, the signal and idler can have all possible combinations of OAM modes with different values of normalized overlap integral. Given that the overlap integral determines the nonlinear gain, under same experimental conditions, different combinations of signal and idler OAM modes, satisfying the OAM conservation, will have different parametric gain and OPG threshold. Using the mathematical treatment[13,20] we have estimated the normalized overlap integral, $\eta_{l_p,l_s,l_i}$, for different combinations of possible OAM modes of signal and idler beams for the pump OAM mode of $l_p$=1. For our experimental conditions, it is interesting to note that $\eta_{l_p,l_s,l_i}$ has highest value of $\eta_{110}$=0.843 for signal and idler OAM mode of $l_s$=1 and $l_i$=0 (Gaussian mode) respectively. On the other hand, for $l_s$=0 (Gaussian) and $l_i$=1, we estimate $\eta_{l_p,l_s,l_i}$ to be $\eta_{1,0,1}$=0.546. Given that the optical beams at longer wavelengths and the higher order OAM modes[21] have higher divergence, the lower value of $\eta_{101}$ for $l_p$= 1, $l_s$= 0 and $l_i$ =1 can be attributed to the higher beam divergence of the idler radiation. Unlike the vortex pumped OPOs where the cavity dictates the beam overlapping[13], here, in single-pass OPG process the intrinsic divergence of the interacting beams determine the effective beam overlapping and thus the transfer of OAM among the interacting beams. The higher nonlinear gain arising from the higher value of $\eta_{110}$=0.843, justifies the signal beam to carry the OAM of the pump beam. Measuring the intensity profile of the signal and idler across tuning range we observed the signal and idler beams to carry vortex and Gaussian spatial structure respectively. Unlike the spontaneous down conversion process[22], it is interesting to note that the signal beam maintains its vortex spatial distribution along beam propagation.

To verify the power scaling characteristics of the OPG source we have recorded the signal power at wavelength, $\lambda_s$=1523 nm as a function of pump power while focusing the pump vortex of order, $l_p$=+1 using a lens, $f$=100 mm. The results are shown in Fig. 3. The signal vortex power (solid circle) increases linear to the pump power at a slope efficiency as high as 92% with a threshold of ~2.8 W. The signal vortex has a maximum output power of 1.6 W corresponding to the pump vortex power of 4.66 W at a single-pass vortex-vortex conversion efficiency of 34.3%. As expected, we observed similar performance (open square) of OPG for the pump beam vortex order, $l_p$= -1. Since the OPG process is commonly studied with Gaussian pump beams[16,23,24], we have compared the performance of vortex pumped OPG source with that of the Gaussian pumped OPG. As evident from Fig. 3, the Gaussian beam pumped OPG (open circle) has operation threshold power of 1.24 W, significantly lower than that of the vortex pumped OPG. The signal power (open circle) increases linearly to the Gaussian pump power with a slope efficiency of 52% resulting a maximum signal power of 1.9 W for 4.66 W of pump power at a single-pass efficiency of ~40.8%. The higher operation threshold and lower single-pass efficiency of vortex pumped OPG as compared to that of Gaussian beam pumping can be attributed to the lower parametric gain for vortex pump beam. The reduction in parametric gain for the vortex beam can be understood as follows. The vortex beam has higher divergence as compared to the Gaussian beam[19] and increases with the order of the beam. Since the beam divergence reduces overlapping among the interacting beams inside the nonlinear crystal, the parametric gain of vortex beam is lower than that of Gaussian beam and reduces with the order of the pump vortex. For this reason, we could not realize OPG for the vortex of order, $l_p=\pm2$. One can in principle produce OPG of higher order vortices with the increase of average pump power, however, crystal damage at elevated pump power can be the major limitation.

To study the performance of the vortex pumped OPG, we pumped the nonlinear crystal at a constant power of ~4.72 W and vortex order of, $l_p=1$, and recorded the signal power, idler power, and corresponding pump depletion across the tuning by varying the crystal temperature and the grating periods[16]. As evident from Fig. 4(a), the signal beam of vortex order, $l_s=1$, can be tuned continuously in the near-IR wavelength range across 1433-1553 nm with a maximum power of 1.7 W at 1509 nm and >0.5 W of output power across 95% of the tuning range. Additionally, the vortex pumped OPG produces idler radiation (see Fig. 4(b)) in Gaussian profile tunable in the mid-IR wavelength range across 3379-4132 nm with the maximum power of 0.48 W at 3625 nm and a usable idler power >0.1 W across 90% of the tuning range. As evident from Fig. 4(c), the vortex pumped OPG has maximum pump depletion of ~71% at the signal wavelength of 1493 nm and >40% of pump depletion across almost the entire tuning range. Since the direction of phase variation of the vortex beam has no effect in the parametric gain of the nonlinear processes, we have observed similar performance of the OPG for pump vortex, $l_p=-1$.

Further, we have studied the spectral characteristics of the pump, signal and idler radiations with the results shown in Fig. 5. As evident from Fig. 5(a), the signal beam of vortex order, $l_s=+1$ has the spectral width (FWHM) of $\Delta\lambda_s$~4.3 nm (~18.5 cm$^{-1}$) centered at $\lambda_s=1523$ nm. On the other hand, the idler beam (see Fig. 5(b)) of Gaussian spatial profile ($l_i=0$) has a measured spectral width (FWHM) of $\Delta\lambda_i$~129.5 nm (~101.5 cm$^{-1}$) centered at $\lambda_i=3572$ nm, clearly showing the generation of broadband idler radiation in the mid-IR wavelength range. Further to understand the broadband generation, we have measured the spectral width of the vortex pump beam ($l_p=+1$) before and after the OPG crystal. As evident from Fig. 5(c), the pump beam (measured before the crystal, black line) having spectral bandwidth (FWHM) of $\Delta\lambda_p$~15 nm (132.5 cm$^{-1}$) centered at $\lambda_p$~1064 nm entirely depleted (as measured after the crystal, red line) due to the presence of broad wavelength acceptance bandwidth of 50-mm long MgO:PPLN crystal. As reported previously[14,15], such a large pump acceptance bandwidth can be attributed to the zero GVM of MgO:PPLN crystal at temperature, T=100°C and period, Λ=30 μm. However, the small discrepancy between the spectral width of pump with the sum of the spectral width of signal and idler can be attributed to the measurement error due to the low resolution (~0.5 nm) of the spectrometer.

We have also studied the temporal and spectral characteristics of the signal and idler radiation of the vortex pumped OPG. Pumping with vortex order, $l_p$=1, at constant power of 4.66 W, we have measured the temporal and spectral width of the signal and the spectral width of the idler across the tuning range. Using the intensity autocorrelation we measured the signal vortex at 1523 nm has a pulse-width (FWHM) of 637 *fs*. Similarly, the temporal and spectral width of the signal radiation vary in the range of 616 *fs* to 544 *fs* and from 4.8 nm to 5.8 nm respectively across 1450 nm to 1544 nm. The calculated time-bandwidth product by assuming the *sech$^2$* pulse shapes varies from 0.42 at 1450 nm to 0.40 at 1544 nm. The spectral width of the idler radiation varies from 38 nm to 120 nm across 3494.5-4029.7 nm with a maximum of spectral width of 129.5 nm at 3572 nm. The signal and idler exhibit a peak-to-peak passive power stability better than 3% and 1.1% respectively over 30 minutes.

In conclusion, using single-pass OPG of ultrafast laser in a 50-mm long PPLN crystal with pump vortex of order, $l_p$=1, we have generated high RR, near-IR signal beam of vortex order, $l_s$=1, tunable across 1433-1553 nm. The signal vortex beam has a maximum power of 1.7 W and the idler power of 0.48 W in Gaussian profile for 4.72 W of pump vortex power corresponds to the single-pass OPG conversion efficiency as high as 46.1%. This is the highest single-pass OPG efficiency for the beams other than Gaussian beams. The signal radiation has the spectral and pulse width of 4.3 nm and 637 *fs* respectively, corresponding to the time-bandwidth product of 0.44 at 1523 nm. In addition, the idler beam has the bandwidth of 129.5 nm at 3572 nm.

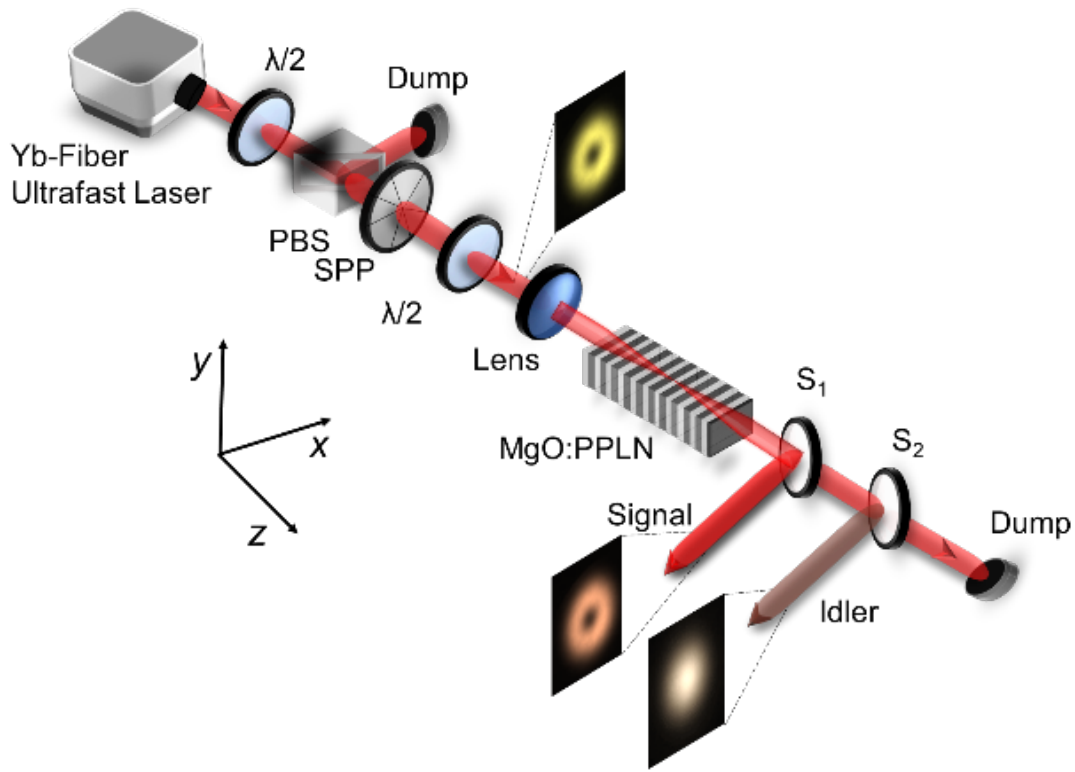

**Figure 1**

Fig. 1. Schematic of the experimental setup for vortex pumped OPG. λ/2, half-wave plate; PBS, polarized beam splitter; SPP, spiral phase plate; MgO:PPLN, nonlinear crystal; S1, S2, wavelength separators. Inset represent the beam profile of the pump, signal and idler radiations.

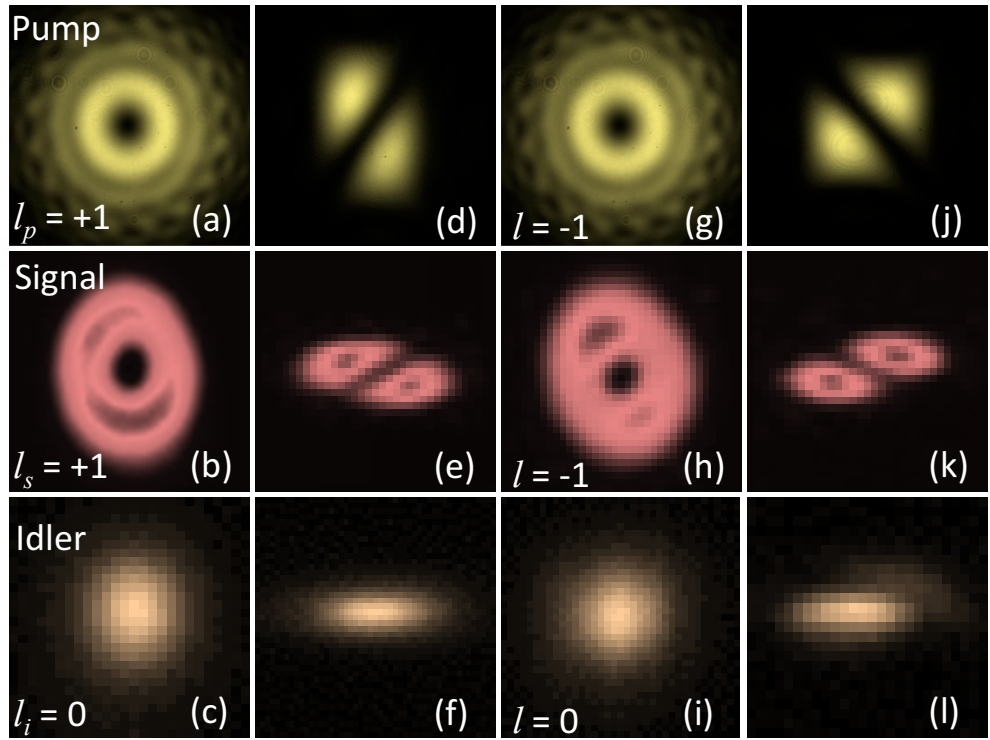

**Figure 2**

Fig. 2. Far-field intensity distribution, (a)-(c), and corresponding characteristic lobes, (d)-(f), of pump, signal and idler for the pump vortex order, $l_p$= 1. Far-field intensity distribution, (g)-(i), and corresponding characteristic lobes, (j)-(l), of pump, signal and idler for the pump vortex order $l_p$= -1.

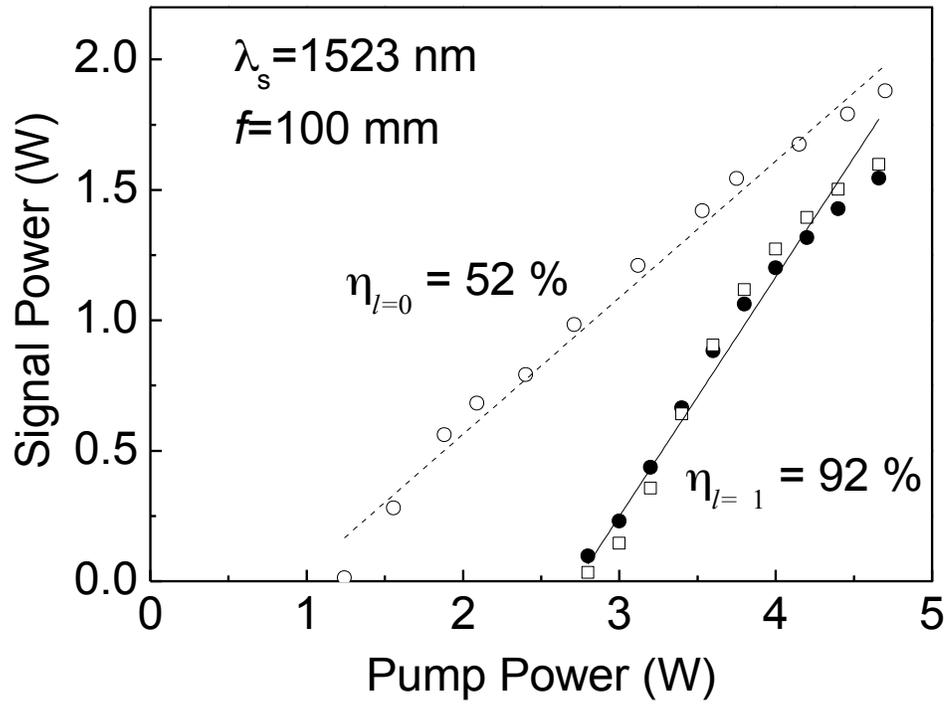

**Figure 3**

Fig. 3. Dependence of signal power on the power of pump beams in vortex and Gaussian profiles. Lines are linear fit to the measured data.

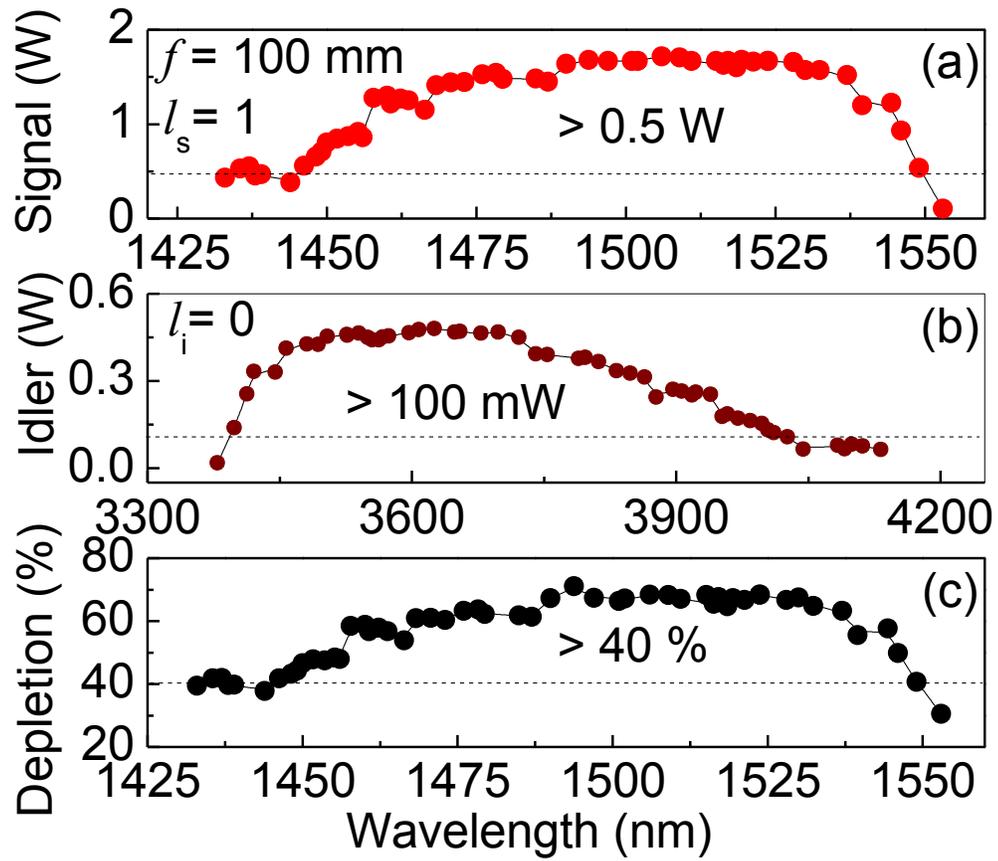

**Figure 4**

Fig. 4. Variation of signal, idler power and pump depletion across the tuning for the pump vortex of order, $l_p = 1$. Pump power is 4.72 W.

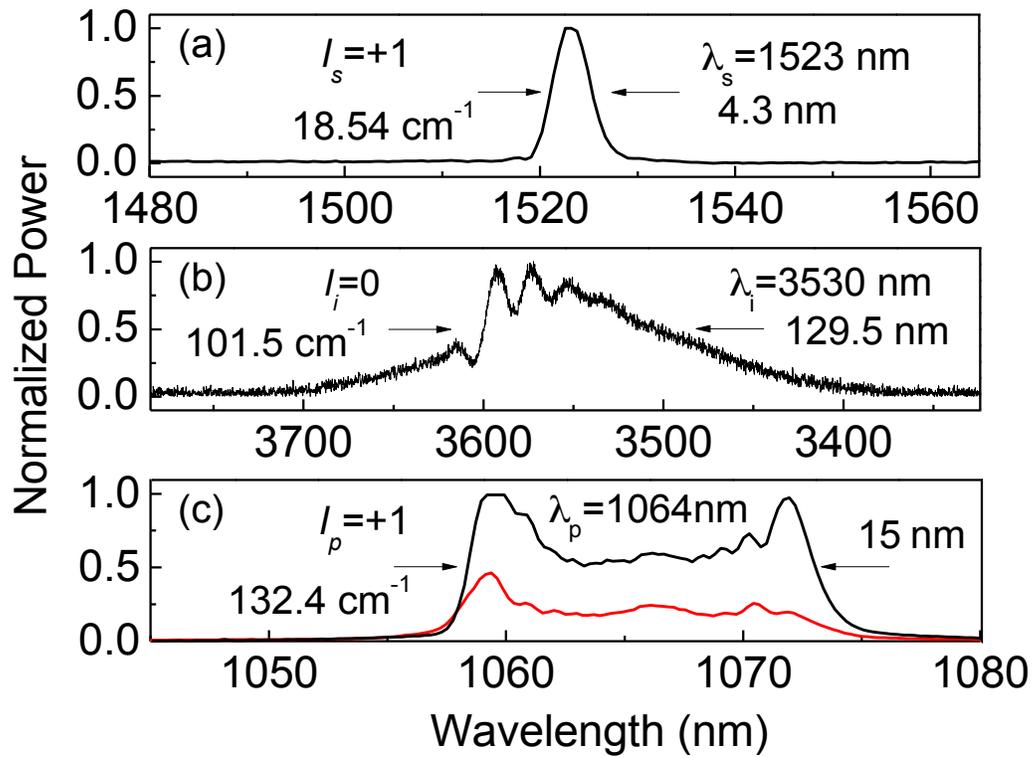

**Figure 5**

Fig. 5. Spectral distribution of the (a) signal, (b) idler, and (c) pump beam.